\documentstyle[twocolumn,prl,aps,epsf,floats]{revtex}
\newcommand{\bea}{\begin{eqnarray}}
\newcommand{\eea}{\end{eqnarray}}
\newcommand{\beq}{\begin{equation}}
\newcommand{\eeq}{\end{equation}}
\newcommand{\la}[1]{\label{#1}}
\newcommand{\doo}{\partial}

\newcommand{\vwall}{v_{\rm wall}}
\newcommand{\vslow}{v_{\rm slow}}

\newcommand{\req}{\hat r_{\rm eq}}
\newcommand{\vali}{\hspace{0.1cm}}
\begin{document}
\draft
\title{Real-Time History of the Cosmological Electroweak Phase Transition}
\author{H. Kurki-Suonio$^{a,}$\cite{mailh}
and M. Laine$^{b,}$\cite{mailm}}
\address{$^a$Department of Physics, 
P.O.Box 9, FIN-00014 University of Helsinki, Finland \\ $^b$Institut
f\"ur Theoretische Physik, Philosophenweg 16, D-69120 Heidelberg,
Germany}
\date{August 1, 1996}
\maketitle

\begin{abstract}
We study numerically the real-time history of the cosmological
electroweak phase transition, as it may take place in the Standard
Model or in MSSM for $m_H\lesssim m_W$ according to recent lattice
results. We follow the nucleated bubbles from the initial stages of
acceleration and rapid growth, through collisions with compression waves
resulting in slowing down and reheating to $T_c$, until the final
stages of slow growth and evaporation. We find that collisions with
compression waves may make the bubble walls oscillate in the radial
direction, and that reheating to $T_c$ takes generically place.
\end{abstract}
\narrowtext

\vspace*{-9.5cm}
\noindent
\hspace*{9.5cm} \mbox{HU-TFT-96-26, HD-THEP-96-22, hep-ph/9607382}
\vspace*{8.3cm}

The electroweak phase transition in the early Universe may have had
important consequences, like baryogenesis~\cite{krs}.  These
depend crucially on the non-equilibrium details of the
first-order transition~\cite{rs}.  The general
features of the real-time history of first-order cosmological phase
transitions have been known for quite a long time~\cite{st}, and, e.g.,
the properties of bubble growth soon after the nucleation period have
been studied in
detail~[4--6]. 
The whole real-time history
has also been computed in some approximation~\cite{rth,H}, but a more 
detailed hydrodynamical investigation has so far been missing. In addition, 
the relevant parameter values are reliably known only after recent
lattice simulations~\cite{klrs}.
The purpose of this paper is to
present a detailed hydrodynamical study of the
complete real-time history of the electroweak phase transition
for realistic parameter values.

Before the transition the universe is in the 
high-temperature `symmetric' phase.  As the
temperature falls below the critical temperature $T_c$, the symmetric phase
becomes metastable.  Bubbles of the `broken-symmetry' 
phase nucleate at $T_n < T_c$, and grow rapidly.  
There are two modes of bubble growth at this
stage: detonations, which are supersonic, and deflagrations, which are
usually subsonic~\cite{IKKL,KL2}. 
In a deflagration the bubble wall is preceded by a
compression wave, whereas in a detonation it is followed
by a rarefaction wave.
When the bubbles collide, the evolution becomes complicated. 
We find that the compression waves from neighboring bubbles may cause
large-amplitude back-and-forth motion of the bubble wall.  Since the baryon
number is generated at the wall in a manner sensitive to the wall velocity,
these details are important for baryon number production.

The tool used here is a hydrodynamical model
where a scalar order parameter $\phi$ drives the
transition~\cite{IKKL}. 
The (non-perturbative)
effective potential $V(\phi)$ for the scalar order parameter is
parametrized by the latent heat $\hat{L}\equiv L/T_c^4$, the surface tension
$\hat{\sigma}\equiv \sigma/T_c^3$ and 
the correlation length $\hat{l}_c\equiv l_cT_c$.
Non-equilibrium phenomena at
the phase transition front are described by a simple friction term,
and the corresponding parameter $\eta$ is deduced from a recent
microscopic calculation~\cite{mp2}.  

The equations of motion follow from energy-momentum conservation and
an equation for entropy production~\cite{IKKL}.  In an expanding
universe with metric
\[
ds^2 = -dt^2 + R(t)^2(dr^2 + r^2d\theta^2 + r^2\sin^2\!\theta\,
d\phi^2)
\]
and in a spherically symmetric case, eqs.~(5--7) of~\cite{KL2} become
\begin{eqnarray}
\partial_t^2\phi &-& \frac{1}{R^2}\frac{1}{r^2}\partial_r(r^2\partial_r\phi)+
3H\doo_t\phi +
\frac{\partial V}{\partial\phi} \nonumber \\
 & = & -\eta\gamma(\partial_t\phi + v\partial_r\phi), \la{phi} \\
\partial_t E &+& \frac{1}{r^2}\partial_r(r^2 Ev) + 3HE \nonumber \\
 & + & p\bigl[\partial_t\gamma + \frac{1}{r^2}\partial_r(r^2\gamma v)
  + 3H\gamma \bigr] \nonumber \\
 & - & \frac{\partial V}{\partial\phi}\gamma (\partial_t\phi +
v\partial_r\phi) =
\eta\gamma^2(\partial_t\phi + v\partial_r\phi)^2, \la{E} \\
\partial_t Z &+& \frac{1}{r^2}\partial_r(r^2 Zv) + 5HZ +
 \frac{1}{R^2}\biggl[\partial_r p + \frac{\partial V}{\partial\phi}
\partial_r\phi\biggr] \nonumber \\
 & = & -\frac{1}{R^2}\eta\gamma(\partial_t\phi +
 v\partial_r\phi)\partial_r\phi. \la{Z}
\end{eqnarray}
Here $\epsilon$ is the energy density and $w=\epsilon+p$ the enthalpy
density, and we have defined $E \equiv \epsilon\gamma$, $Z \equiv
w\gamma^2v$.  The velocity $v$ above is the coordinate velocity $dr/dt$ of
the fluid, the physical velocity being $Rv$.
The scale factor $R$ changes very little during the period of interest, so the
difference is unimportant, but below $v$ will refer to the physical velocity.
The main effect of the expansion is the cooling of the universe, where the
small change is significant because we are very close to
$T_c$. 
$H = \dot R/R$ is the expansion rate of the universe.

For the thermodynamical properties of the phase transition 
we use four sets of parameters, suggested by lattice simulations~\cite{klrs}.
The main difference with respect to perturbation theory is that 
the surface tension $\hat{\sigma}$ gets small if the transition 
is weak. As the first set we take 
the Standard Model with an unrealistically small Higgs mass
$m_H = 51$ GeV (case C):
\beq
\hat{L} = 0.124,\quad \hat{\sigma} = 0.0023,\quad \hat{l}_c = 8.
\eeq
A second set is $m_H = 68$ GeV in which case the transition
is very weak but still of first order (case D):
\beq
\hat{L} = 0.08,\quad \hat{\sigma} = 0.0002,\quad \hat{l}_c = 10.
\eeq
Both of these cases lead to a transition too weak for 
baryogenesis. As a third case we consider a stronger 
transition (case B):
\beq
\hat{L} = 0.3,\quad \hat{\sigma} = 0.01, \quad \hat{l}_c = 5.
\la{B}
\eeq
This transition corresponds to a tree-level Higgs mass
parameter $m_H^*\sim 50$ GeV and is close to being
strong enough for baryogenesis~\cite{klrs} (we do not consider
the possibilities proposed in~\cite{rate}). 
In addition, the values in eq.~(\ref{B})
could be achieved in MSSM
for a realistic pole Higgs mass $m_H \lesssim m_W$~\cite{mssm}. 
A still stronger transition with $m_H^*\sim 45$ GeV would give
$\hat{L}\sim 0.5$, $\hat{\sigma}\sim0.02$ in MSSM. However, 
to display more clearly the parameter dependence, we 
take rather eq.~(\ref{B}) with just
a larger surface tension $\hat{\sigma} = 0.02$ (case A).

The dependence of the results 
on the parameters is roughly
the following. The supercooling $1-\hat{T}_n$ ($\hat{T}_n\equiv
T_n/T_c$) and the distances $l_n/t_H$ of the nucleated bubbles 
($t_H \equiv H^{-1}$ is the Hubble time, or Hubble length)  
are proportional
to $\hat{\sigma}^{3/2}/\hat{L}$. The velocity of the 
bubble wall during the
rapid growth is proportional to $\hat{\sigma}^{1/2}$. 
Reheating to $T_c$ takes place if 
$\hat{L} \gtrsim 8\hat{\sigma}^{3/4}$, and the velocity
during the period of slow growth is proportional to 
$\hat{\sigma}^{3/2}/\hat{L}^2$, provided that this
number is small enough. 
Hence a 
small $\hat{\sigma}$ indicated by lattice simulations 
tends to make supercooling and velocities small
and reheating likely. 

The friction parameter $\eta$ is determined from the microscopic
analysis in~\cite{mp2}. 
In general, the friction term is non-local and the coefficient
$\eta$ depends
on $\phi$~\cite{thankmoore}.  However, 
we do not here study the microscopic structure of
the phase transition front and hence $\eta$ is only used 
in parametrizing the entropy production.  
Consequently, we may use a simpler
$\phi$-independent (but $m_H$-dependent) $\eta$.  After determining
$\hat{L}, \hat{\sigma}, \hat{l}_c$ 
and the nucleation temperature $\hat{T}_n$
corresponding to the effective potential used
in~\cite{mp2}, we search for
the value of $\eta$ producing the same velocity as in Table~1 of~\cite{mp2}.
We find that $\eta\sim(0.04 - 0.1)T_c$ for the Higgs masses
studied. For completeness we also inspect more
general values of $\eta$ below.

\begin{table}[t]
\squeezetable
\centering
 
\begin{tabular}{|l|l|l|@{}l|l|l|l|@{}l|}
\hline 
case &  $\hat{T}_n$ \mbox{ } & $r_n/t_H$ & \vali
$\eta/T_c$ & $\vwall$    
& $\hat{r}_{\rm eq}$ & $\vslow$ & \vali $\Delta t/t_H$  \\ \hline
\mbox{} A0 & 0.994938 & 2.59$\times10^{-5}$ & \vali 0.003 & 0.959 & 
& & \vali 2.7$\times10^{-5}$
  \\ \cline{1-1} \cline{4-5} \cline{8-8} 
\mbox{} A1 & & & \vali 0.01 & 0.506 &
& & \vali 6.7$\times10^{-5}$
  \\ \cline{1-1} \cline{4-5} \cline{8-8} 
\mbox{} A2 & & & \vali 0.03 & 0.490 &
& & \vali 6.0$\times10^{-5}$
  \\ \cline{1-1} \cline{4-5} \cline{8-8} 
\mbox{} A3 & & & \vali 0.10 & 0.402 &
& & \vali 7.0$\times10^{-5}$
  \\ \hline
\mbox{} B1 & 0.998295 & 8.73$\times10^{-6}$ & \vali 0.01 & 0.406 & 
0.936 & 0.00149 & \vali 3.7$\times10^{-4}$
  \\ \cline{1-1} \cline{4-5}  
\mbox{} B2 & & & \vali 0.03 & 0.379 &
& &
  \\ \cline{1-1} \cline{4-5} 
\mbox{} B3 & & & \vali 0.10 & 0.281 &
& &
  \\ \hline
\mbox{} C1 & 0.999552 & 2.35$\times10^{-6}$ & \vali 0.01 & 0.335 & 
0.796 & 0.00109 & \vali 4.4$\times10^{-4}$
  \\ \cline{1-1} \cline{4-5} 
\mbox{} C2 & & & \vali 0.03 & 0.288 &
& &
  \\ \cline{1-1} \cline{4-5} 
\mbox{} C3 & & & \vali 0.10 & 0.170 &
& &
  \\ \hline
\mbox{} D0 & 0.9999828 & 9.54$\times10^{-8}$ & \vali 0.001 & 0.089 & 
0.311 & 0.00012 & \vali 5.6$\times10^{-4}$
  \\ \cline{1-1} \cline{4-5} 
\mbox{} D1 & & & \vali 0.01 & 0.084 &
& &
  \\ \cline{1-1} \cline{4-5} 
\mbox{} D3 & & & \vali 0.10 & 0.046 &
& &
  \\ \hline
\end{tabular}
\vspace*{3mm}

\caption[a]{
The properties of the transition.  
In case A there is no stage of slow growth, hence no $\hat{r}_{\rm eq}$, 
$\vslow$ are given.  }
\vspace*{-3mm}
\end{table}

For our four parameter sets A--D, 
we determine $\hat{T}_n$ 
and the distances $l_n/t_H$ between the nucleated bubbles~\cite{ghe}
with the following simple procedure, giving better
accuracy than the classical calculation. 
The parameters $\hat{L}$, $\hat{\sigma}$, $\hat{l}_c$ 
fix uniquely a quartic potential $V(\phi)$, 
and for this potential the nucleation action is known 
analytically with good accuracy~\cite{j}.
We then solve for $\hat{T}_n$ 
and $l_n/t_H$  
numerically from eqs.~(4.5), (4.8) in~\cite{eikr}.
By comparing with nucleation calculations 
for more complicated functional forms of $V(\phi)$, 
we have found that
$\hat{l}_c$ should be interpreted as the correlation 
length in the broken phase.

After the nucleation the bubbles start to grow.
We have studied the bubble growth with a spherically symmetric hydrodynamical
code, which solves eqs.~(\ref{phi}--\ref{Z}).
We present results for 13
different sets of parameter values (cases A--D with different values of
$\eta$), see Table~I.
  
\begin{figure}[tbh]
\vspace*{-3.0cm}
\hspace*{-1.2cm}
\epsfysize=14.0cm
\epsffile{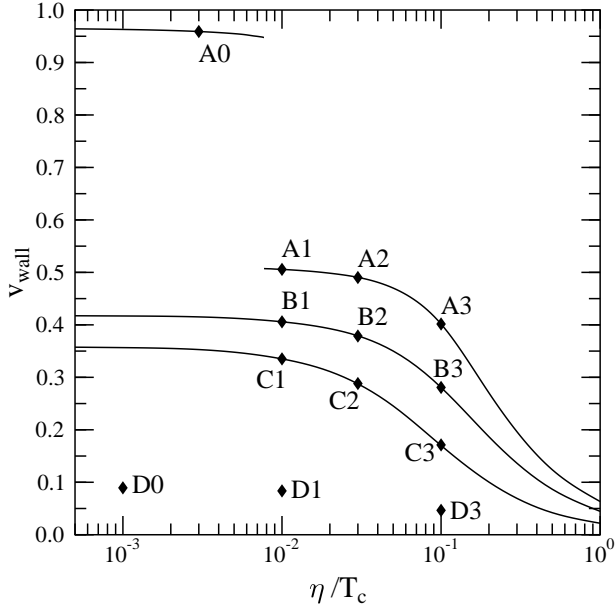}

\vspace*{-2.8cm}

\caption[a]{\protect
The wall velocity as a function of $\eta$.  In case A the wall is a detonation
front for $\eta \lesssim 0.0077 T_c$.  Otherwise the wall is a deflagration 
front.  The black diamonds correspond to the 13 runs discussed in the text. }
\label{eta}
\end{figure}

The initial stages of bubble growth are not affected by the expansion
of the universe or
the existence of neighboring bubbles.  The bubble wall accelerates
and reaches a stationary velocity $\vwall$.  The velocity depends on
the friction at the wall, parametrized by $\eta$. If the
supercooling is relatively large, 
$\vwall$ covers most of the range from 0 to 1
 as $\eta$ is
varied, and the configuration goes through the different modes
described in~\cite{IKKL}. 
For small supercooling, only deflagrations are possible and there is a
maximum wall velocity $v_{\rm max}$\cite{KS}.  We plot $\vwall(\eta)$
in Fig.~\ref{eta}.  
Thus in case D the bubble will be a slowly growing deflagration
bubble. In cases B and C the bubble has to be a deflagration, but
velocities closer to the speed of sound could be achieved.  In case A
detonations are possible.  Case D takes longer to compute, so we show
$\vwall$ for 3 values of $\eta$ only.

After the period of stationary growth, the bubbles
collide with the compression waves from other bubbles.
Our grid length corresponds to  $r_n \equiv l_n/2$,
and we use reflective boundary
conditions at the (spherical) outer edge to represent collisions with the
neighboring bubbles.
This is an unrealistic geometry, since it makes the collision
simultaneous at all points of the bubble wall, preserving the spherical
symmetry.  
Since the nucleation process places the bubbles randomly,
in reality every bubble
has a different collision geometry.  The results presented here give a
qualitative idea of the real events, although the aspherical features and
transverse motions are missing.

\begin{figure}[tbh]
\vspace*{-3.0cm}
\hspace*{-1.2cm}
\epsfysize=14.0cm
\epsffile{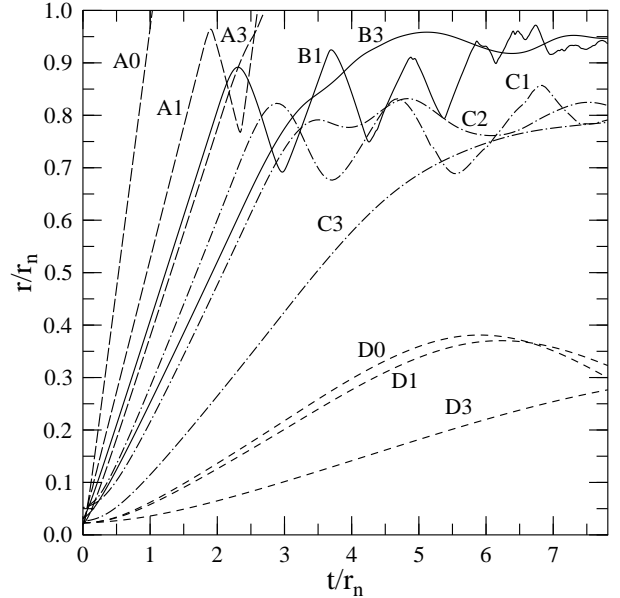}

\vspace*{-2.8cm}

\caption[a]{\protect
The bubble radius as a function of time. For clarity the cases 
A2 and B2 
are omitted.}
\label{xwallm}
\end{figure}

The bubble radius $\hat r \equiv r/r_n$ as a function of time
for 11 of the 13 cases is shown in
Fig.~\ref{xwallm}.  A fraction $f_V = \hat r^3$ of the volume
is in the broken phase (bubble) 
and the transition is completed when $\hat r = 1$.
The microscopic (bubble nucleation radius) and the macroscopic (distance
between bubbles) scales in the problem differ by some 9 orders
of magnitude.
For practical reasons we have set the scales closer to each other on the
computer.  Therefore the duration of the initial acceleration stage appears
greatly magnified in the figure.  

We discuss the fastest-growing bubbles first.  In case A0 the bubble wall is a
detonation front.  No information is transmitted ahead of a detonation, and
therefore the wall propagates undisturbed until it collides with
the neighboring bubble wall.  The phase transition is completed and all the
resulting fluid motion takes place in the broken phase.

In all the other cases the bubble wall is a deflagration front, preceded by a
compression wave in the symmetric phase.  The bubble wall slows down as it
meets the compression wave from the neighboring bubble.  In most cases studied
the compression wave is strong enough to stop the wall and push it back,
compressing the bubble.  As the compression waves move back and forth between
the bubbles, the bubble volume oscillates.  In a realistic geometry the
bubble is deformed as different parts of the bubble wall move with
different velocities.  The typical amplitude of the motion should, however, be
the same as obtained here.  Part of the compression wave is
transmitted through the wall into the broken phase, causing partial
superheating.
In some cases a region inside the bubble may be heated so much that the broken
phase cannot exist at this temperature, and a droplet of the symmetric phase
is temporarily formed. 

Further history of the bubble depends on the magnitude
of the latent heat relative to the
degree of supercooling.  If the latent heat is large enough to reheat the
universe back to the critical temperature, both phases can coexist at $T_c$.
The initial volume fraction of the broken phase at equilibrium 
(ignoring the expansion of the universe since bubble nucleation)
is then $f_V \equiv \req^3 = \epsilon_c(1-\hat{T}_n^4)/L$,
where $\epsilon_c$ is the energy density of the symmetric phase at $T_c$.
The bubble radius oscillates
around $\req$.  The initial amplitude
depends on $\vwall$, as deflagrations close to
the speed of sound are preceded by strong compression waves 
which cause large oscillations.
A smaller $\req$ also leads to larger 
oscillations as the compression waves travel longer between the bubbles.
The oscillation amplitude decreases with time and the bubble settles to a
long period of slow growth.

In case A the nucleation temperature is so low that 
the universe does not reheat to $T_c$.
The transition is completed after at most one oscillation.
The duration of the phase transition $\Delta t$
is given essentially by the bubble
separation and the wall velocity, $\Delta t = r_n/\vwall$.
In case A1 the wall changes nature from a deflagration to a detonation after
the collision.

In cases B, C, and D, 
the universe reheats to $T_c$
and the bubble radius oscillates around the equilibrium value $\req$, 
given in Table I.

\begin{figure}[t] 
\vspace*{-3.0cm}
\hspace*{-1.2cm}
\epsfysize=14.0cm
\epsffile{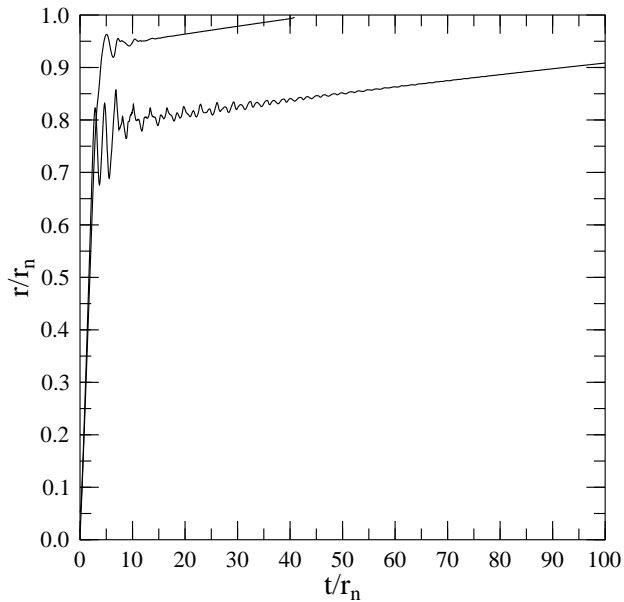}

\vspace*{-2.8cm}

\caption[a]{\protect
The bubble history showing the stages of fast and slow growth.  
The upper line is case B3 and the lower one C1. }
\label{xslow}
\end{figure}

The period of slow growth is paced by the expansion of the universe, which
slowly makes space for the released latent heat. This takes a much longer time
than the preceding stages, see Fig.~\ref{xslow}. To first order in 
$L/\epsilon_c$, the duration of the phase
transition is
$  
\Delta t/t_H=L(1-\req^3)/(4\epsilon_c),
$  
assuming that in the symmetric phase $p\propto T^4$. 
The wall velocity $\vslow$
during the slow growth is geometry dependent, but for the spherical
bubble geometry we
can estimate it by
$   
 \vslow \sim (1-\req)(r_n/\Delta t).
$  
The values of $\Delta t/t_H$ and $\vslow$ are
displayed in Table I.  
For case D $\vslow$ is much less than for the other cases,
because the bubbles are much closer to each other.

At some point during the slow growth the remaining regions in the
symmetric phase pull themselves to spherical droplets, resulting in a geometry
inverse to the one used above.  We have done runs with an inverted geometry,
i.e., the shrinking symmetric phase at the center surrounded by the broken
phase, starting at a static equilibrium situation at $T \sim T_c$.  The
droplet evolves first towards the weak deflagration similarity
solution described in~\cite{RMP} 
and the end stages are as found in~\cite{KL2}.

The phenomena discussed might have physical significance, e.g., 
for baryon number generation. Indeed, the velocity dependence
of baryon number production may be such that the slowing
down of $\vwall$ through collisions with compression
waves, and the resulting oscillations, have a favourable effect on 
the baryon number produced~\cite{H,rs}.
It should be noted, however, that for stronger transitions
(D$\to$B) where the baryon number is not washed out afterwards, 
the fraction of space where the transition
proceeds through slow growth is smaller. 

Finally, let us note that there remain
some important open questions 
concerning the real-time history. 
In particular, while the bubble wall appears to be stable
during the period of fast growth~\cite{hkllm}, the period of 
slow growth is different in character and the stability 
properties remain open. A related problem is the 
appearance of turbulence, which might 
affect, e.g., magnetic fields~\cite{bbl}.

We thank J. Ignatius, K. Kajantie, G. D. Moore and 
M. Shaposhnikov for discussions and 
the Center for Scientific Computing (Finland) for
computational resources.  M.L was partially supported by the
University of Helsinki.


\begin{thebibliography}{99}
\bibitem{ghe}
'Subcritical bubbles' (e.g.~\cite{gh})
refer to the breakdown of perturbation theory at large $m_H$, 
and are of no concern here since non-perturbative lattice results
are used.
\end{thebibliography}
\end{document}